# Unravelling the role of Sm $4f$ electrons in the magnetism of SmFeO$_3$

Danila Amoroso,[1] Bertrand Dupé,[2,1] and Matthieu J. Verstraete[3,*]

[1]*Nanomat/Q-mat/CESAM, Université de Liège, B-4000 Sart Tilman, Belgium*
[2]*Fonds de la Recherche Scientifique (FNRS-FRS), B-1000 Brussels, Belgium*
[3]*Nanomat/Q-mat/CESAM, Université de Liège, and European Theoretical Spectroscopy Facility, B-4000 Sart Tilman, Belgium*

Magnetic rare-earth orthoferrites $R$FeO$_3$ host a variety of functional properties from multiferroicity and strong magnetostriction, to spin-reorientation transitions and ultrafast light-driven manipulation of magnetism, which can be exploited in spintronics and next-generation devices. Among these systems, SmFeO$_3$ is attracting a particular interest for its rich phase diagram and the high temperature Fe-spin magnetic transitions, which combines with a very low temperature and as yet unclear Sm-spin ordering. Various experiments suggest that the interaction between the Sm and Fe magnetic moments (further supported by the magnetic anisotropy), is at the origin of the complex cascade of transitions, but a conclusive and clear picture has not yet been reached. In this work, by means of comprehensive first-principles calculations, we unravel the role of the magnetic Sm ions in the Fe-spin reorientation transition and in the detected anomalies in the lattice vibrational spectrum, which are a signature of a relevant spin-phonon coupling. By including both Sm-$f$ electrons and non-collinear magnetism, we find frustrated and anisotropic Sm interactions, and a large magnetocrystalline anisotropy mediated by the SOC of the Sm-$4f$ electrons, which drive the complex magnetic properties and phase diagram of SmFeO$_3$.

## I. INTRODUCTION

Samarium ferrite, SmFeO$_3$ (SFO), belongs to the family of magnetic perovskite oxides of the type $RM$O$_3$, with $R=$ rare-earth and $M=$ Cr, Mn or Fe (see Ref. [1] and references therein). These compounds host two different magnetic sub-lattices at the $A$ and $B$ sites of the $AB$O$_3$ perovskite structure, giving rise to competing magnetic interactions. This is of fundamental interest for the physics of magnetism, and also has potential for applications in functional materials and spintronics [2]. Two central phenomena are observed in these perovskite systems: a temperature-dependent spin reorientation (SR) process and/or a magnetization compensation and reversal, which shows the importance of the entropy contribution in the stabilization of magnetic state. In magnetic materials, heat is absorbed in magnonic excitations: the presence of rare-earth elements with high magnetic moments opens up the possibility to reach large magnetoelectric effect [3].

Among these compounds, SFO undergoes various phase transitions and shows fast magnetic switching [4–6]: at the Néel temperature (T$_N \simeq$ 670 K), iron spins order antiferromagnetically along the ***a***-axis of the $Pbnm$ crystal structure, with an additional, weak ferromagnetic (wFM) component along the ***c***-axis due to the spin canting; between 450 K and 480 K (T$_{SR}$) a rotation of the easy axis associated to the wFM takes place, with a reorientation of the iron spins from the ***a***-axis to the ***c***-axis; below 100 K the net magnetization associated to the canted spins monotonically decreases and reverses sign at very low temperature, passing through a compensation point (T$_{comp} \simeq$ 4 K) with zero magnetization. Such a behavior at low temperature is interpreted as the appearance of weak ferromagnetism associated to the Sm-spin sub-lattice, which would be opposite in direction with respect to the Fe-wFM. Some experimental attempts to determine SFO magnetization at low temperature are reported, but a precise characterization of the Sm-spin properties remains lacking. Difficulties include the poor stability of the expected Sm-ordering and the strong Sm absorption cross section for neutrons [7–9]. A so-called cluster-glass state (a spin-glass behaviour associated to a magnetostatic excitation through a spin-phonon or a spin-lattice interaction) is also reported in the temperature region between 100 K and 200 K, where the net magnetization reaches its saturation [10]. Because of its high T$_{SR}$ and its rich phase diagram, SFO has attracted particular attention for exploitation in technological devices, and for property engineering ranging from magnetization and ferroelectricity enhancement by strain or oxygen vacancies [11, 12], to nanostructure fabrication [13, 14] and Sm-site doping [15, 16]. Interestingly, SFO has been also investigated for gas sensing [17, 18], with a recent proposal of an application for non-invasive colorectal cancer screening [19].

Similarly to the case of other $R$FeO$_3$ systems, such as NdFeO$_3$ [20, 21] or TmFeO$_3$ [22], the SR transition in SFO is considered to be strongly dependent on the magnetic anisotropy related to the $R$-4$f$ electrons. Nevertheless, theoretical studies aimed at understanding the microscopic mechanisms driving the spin reorientation transition in SFO remain limited due to the structural, electronic, and magnetic complexity of the material [23–28]. In particular, the contribution of Sm 4$f$ electrons and spin-orbit coupling (SOC) has often been neglected in first-principles calculations, due to the complexity of treating $f$ states and non-collinear magnetism.

In this work, we report an investigation of the crystal and magnetic structures of SFO, exploiting density functional theory (DFT) simulations and accounting explicitly for the Sm-magnetism through the inclusion of the Sm-$4f$ electrons in the valence states. First, we show that Sm-$f$ electrons are essential for the good description of the crystal structure (lattice parameters and interatomic distances) by comparison to available experimental data. Then, we provide estimates of the magnetic interactions in terms of effective Heisenberg Fe-Fe, Sm-Sm and Fe-Sm exchange couplings, and of the magnetic anisotropy by taking into account the SOC effect, and associated non-collinear spin states for the Fe- and Sm-substructures. Finally, we report the results of phonon calculations with a particular focus on the Raman active modes in the $Pbnm$ crystal structure. Our results support the hypothesis of an active role for Sm magnetism in the experimentally observed anomalies in the Raman spectrum evolution as a function of temperature [29].

---

* Matthieu.Verstraete@uliege.be

To the best of our knowledge this is the first time a magnetic parametrization and magnetic anisotropy estimate have been carried out for SFO. We find that it is the strong magnetic anisotropy of the Sm $4f$ electrons which drives the high-temperature Fe spin reorientation, mediated by a small effective Sm-Fe magnetic exchange.

## II. COMPUTATIONAL DETAILS

DFT calculations were performed using the projector augmented wave (PAW) method[30] as implemented in the ABINIT simulation package [31–34]. The exchange-correlation potential was evaluated within the generalized gradient approximation (GGA) using the PBEsol functional [35]. The following orbitals were considered as valence states: O $2s^2$, $2p^4$; Fe $3s^2$, $3p^6$, $3d^7$, $4s^1$; Sm $5s^2$, $5p^6$, $5d^1$, $6s^2$ without Sm-$f$ electrons (noted w/o Sm-$f$) when treating the Sm-$f$ electrons as core states, and with Sm-$f$ electrons (noted w/ Sm-$f$) when taking Sm-$4f^5$ electrons in the valence. The plane wave cutoff energy was set to 35 Ha and the $k$-mesh for the Brillouin zone (BZ) sampling of the $Pbnm$ 20-atom cell to $6 \times 6 \times 4$.

The self-consistent energy was converged below $10^{-10}$ eV for collinear calculations, without spin-orbit coupling (SOC), and between $10^{-4}$ and $10^{-6}$ eV for calculations including SOC and non-collinear magnetism. No symmetry constraints were taken into account (nsym=1). We fixed the Hubbard-$U$[36, 37] correction on the localized Fe-$3d$ and Sm-$4f$ states to 4.5 eV and 6 eV, respectively. Different $U$ values (5 or 7 eV on Sm-$f$ or 4 eV on Fe-$d$ states) were explored and do not substantially affect the estimate of the effective magnetic interactions, as commented along the text.

The atomic structure and lattice parameters of the orthorhombic 20-atom SFO-cell were fully optimized for both the w/o Sm-$f$ and the w/ Sm-$f$ cases: in the w/o Sm-$f$ case, we employed $U(\text{Fe-}d) = 4.5$ eV and collinear antiferromagnetic (AFM) G-type Fe-spin order; in the w/ Sm-$f$ case, we employed $U(\text{Sm-}f)=6$ and $U(\text{Fe-}d)$ =4.5 eV and AFM G-type order for both the Sm and Fe spin substructures. In the latter case, small Sm and Fe polar distortions along the **c**-crystallographic axis and additional antipolar Fe displacements along the **b**-crystallographic axis appear, lowering the crystal symmetry to the $C_{2v}$-$Pbn2_1$ (n. 33) space group. Nevertheless, distortions with respect to the associated centrosymmetric $D_{2h}$-$Pbnm$ (n. 62) structure were lower than 0.005 Å, and produce an energy gain lower than 0.1 meV/f.u. Given typical DFT accuracies, we can not assert the low energy phases are truly stable, and we decided to work in the SFO-$Pbnm$ phase. Additionally, phonon frequencies calculated at the $\Gamma$-point of the $Pbnm$-BZ through the "finite difference" approach using the Phonopy package [38], do not reveal structural instabilities and are almost the same (differences below 1 cm$^{-1}$) as those calculated in the $Pbn2_1$ structure. We also validated the vibrational energies and phonon eigenvectors of the finite difference method by comparison with density functional perturbation theory (DFPT)[39, 40].

In this work, all results concern calculations performed using the $Pbnm$ structures for both the w/o Sm-$f$ and the w/ Sm-$f$ cases, unless stated otherwise.

## III. CRYSTAL AND ELECTRONIC STRUCTURES

### A. Structural properties

SFO crystallizes in the common $D_{2h} - Pbnm$ (or $Pnma$) orthorhombic perovskite structure [44], which hosts two interpenetrating pseudo-cubic Sm- and Fe-substructures. The crystal structure is depicted in Fig. 1. The main structural distortions from the ideal $Pm\bar{3}m$-cubic cell are those related to the Sm-O bonds: the antipolar off-centering of the samarium $A$-cations and the oxygen atom rotations ($a^-a^-c^+$ in Glazer's notation [42]) produce distorted SmO$_{12}$ dodecahedra and tilted, corner-sharing FeO$_6$ octahedra, respectively.

The calculated structural parameters, including unit cell lattice vectors and angles, interatomic distances, $b/a$ and $c/a$ ratios were calculated and compared to experimental values, as shown in Table I and Table SI. The effect of the Sm-$f$ electrons can already be seen in the structural parameters: the $a$ lattice constant, in particular, increases leading to a bigger volume of the w/ Sm-$f$ unit cell compared to the w/o Sm-$f$ one. Overall structural parameters are very close and in good agreement with available experimental data (T$\sim$ 300 K in Refs. [41, 45] and T$\sim$ 100 K in Ref. [29]). Particularly, by analysing results from structural optimizations performed either in the w/o Sm-$f$ case (freezing Sm-$f$ electrons in the core states) or in the w/ Sm-$f$ case (including them in the valence), we found that in the latter case the agreement with X-ray diffraction (XRD) measurements improves significantly. The relative error (RE) associated to our DFT results with respect to the experimental data is overall reduced when taking into account the $f$-electrons (see RE reported in Tables I-SI).

| Parameters | w/o Sm-$f$ | w/ Sm-$f$ | w/ Sm-$f$ (1 GPa) | Exp. 293 K[41] | RE w/o Sm-$f$ | RE w/ Sm-$f$ | RE w/ Sm-$f$ (1 GPa) |
|---|---|---|---|---|---|---|---|
| $\|a\|$ (Å) | 5.36 | 5.41 | 5.40 | 5.40 | -0.74% | +0.17% | -0.01% |
| $\|b\|$ (Å) | 5.62 | 5.60 | 5.59 | 5.60 | +0.34% | -0.06% | -0.24 % |
| $\|c\|$ (Å) | 7.69 | 7.70 | 7.69 | 7.71 | -0.21% | -0.08% | -0.27 % |
| $\|b/a\|$ | 1.05 | 1.04 | 1.04 | 1.04 | +1.09% | -0.23% | -0.23% |
| $\|c/a\|$ | 1.44 | 1.42 | 1.42 | 1.43 | +0.54% | -0.25% | -0.26% |
| volume (Å$^3$) | 231.59 | 233.07 | 231.80 | 233.01 | -0.61% | +0.03% | -0.52% |

TABLE I. Calculated lattice parameters for orthorhombic $Pbnm$ SFO either without (w/o Sm-$f$) or with (w/ Sm-$f$) the Sm-$f$ electrons as valence states. Room temperature experimental data from Ref. [41] and relative errors (RE) are also reported for direct comparison.

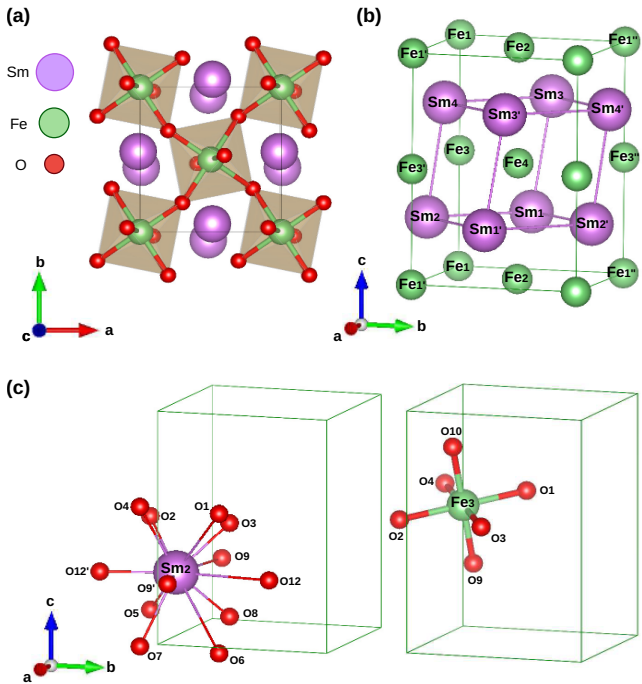

FIG. 1. Orthorhombic SFO structure. **(a)** Top view of the $Pbnm$ structure, showing in-phase oxygen octahedra rotations around the $c$-axis ($c^+$ in Glazer's notation [42]). Samarium (Sm), Iron (Fe) and Oxygen (O) atoms are represented with violet, green and red balls, respectively. **(b)** Sm and Fe atomic arrangement, forming two distorted, pseudo-cubic, magnetic substructures. **(c)** Sm-O (left) and Fe-O (right) bonds, forming distorted SmO$_{12}$ polyhedra and FeO$_6$ octahedra, respectively. Atomic labels guide identification of the different interatomic distances reported in Table SI. Atomic structure were generated through the VESTA software [43].

The structural distortions of the ground state (shown in Fig. 1) produce inequivalent bonds and different distances between neighboring magnetic cations, which creates inequivalent magnetic sites and could have a consequence in the extraction of shell-dependent magnetic interactions. The six nearest neighbor Fe-Fe distances range between 3.85 and 3.89 Å, with the shorter Fe$_1$-Fe$_3$ distance along the $c$ crystallographic vector and the longer distance for the in-plane Fe$_1$-Fe$_2$ pairs ($a, b$ plane). As highlighted in Fig. 1(b) and Fig. S5, the Sm pseudo-cubic cage is more distorted than the Fe one, and the six nearest neighbor Sm-Sm distances range from 3.80 to 3.99 Å. Both Sm-Sm and Fe-Fe pairs count twelve next-nearest (NN) neighbors, with distances ∈ [5.19 : 5.81] Å and [5.41 : 5.60] Å, respectively. Values are reported in Table SI.

We note that the length of the $c$ lattice vector is fixed by the Fe$_1$-Fe$_3$ distance, whereas the length of the planar $a$ and $b$ vectors is determined by the Sm$_1$-Sm$_{1'}$ = Fe$_1$-Fe$_{1'}$ and Sm$_2$-Sm$_{2'}$ = Fe$_1$-Fe$_{1''}$ distances, respectively (*cfr.* Fig. 1(b) and Table SI). Accordingly, one of the effects of the Sm-$f$ electrons on the structure is to introduce some in-plane anisotropic strain: the length of the $a$ lattice vector is significantly reduced w/o Sm-$f$ electrons, whereas $b$ slightly increases; $c$ is less affected (as shown in Table I and Table SI). Similarly, Sm-$f$-related structural anisotropy, mostly affecting the $a$ lattice strain, is also reported for DyFeO$_3$ [46], which suggests that this trend is general. On the other hand, the overall volume w/ Sm-$f$ agrees with that at room temperature although our DFT calculation is in principle done at 0 K. Due to thermal expansion, a reduction at low-T is expected instead [29, 45]. Therefore, to disentangle the effects of isotropic volume compression and anisotropic strain in the dynamical properties reported in Sec. V, the $Pbnm$ w/ Sm-$f$ structure was also relaxed by applying a hydrostatic pressure of about 1 GPa, bringing the volume closer to that w/o Sm-$f$ and also closer to the expected low temperature one.

### B. Electronic properties

In Fig. 2, we show the (collinear spin) electronic partial density of states (DOS) for antiferromagnetic SFO, in the **G**-type spin order, which is characterized by all first-neighbor spins coupled antiferromagnetically (Fig.S4e). Spin-up and spin-down DOS channels associated to one atom of each species are shown. The valence bands (VB) can be divided into four main energy regions: $(-7 : -6)$ eV with prominent Fe-3$d$ states and a small O-$2p$ state contribution; $(-6 : -4.5)$ eV with hybridizing Fe-$4s$ and O-$2p$ states; $(-4.5 : -3)$ eV with prominent localized Sm-$4f$ states, overlapping with small Fe-$3d$ and O-$2p$ states contributions; $(-3 : 0)$ eV (top of VB) characterized by hybridized Fe-$4s$ and O-$2p$ states. The bottom of the conduction bands (CB) is characterized by the remaining empty Sm-$4f$ spin-up states and the empty Fe-$3d$ spin-down states. In Figs. S1-S2, we show the total electronic DOS and band structure calculated in the $Pbnm$ structure w and w/o Sm-$f$. The direct energy band gap at the $X$-point (0.5, 0.0, 0.0) of the Brillouin zone (BZ) is about 2.3 eV within DFT PBEsol+U. The Fe and O electronic states are not significantly affected by the Sm states. In Fig. S2, we compare the band structures obtained by including and excluding the Sm-$4f$ electrons from the valence states. The main effects are in the middle of the VB and bottom of the CB, related to the appearance of the localized $f$-states.

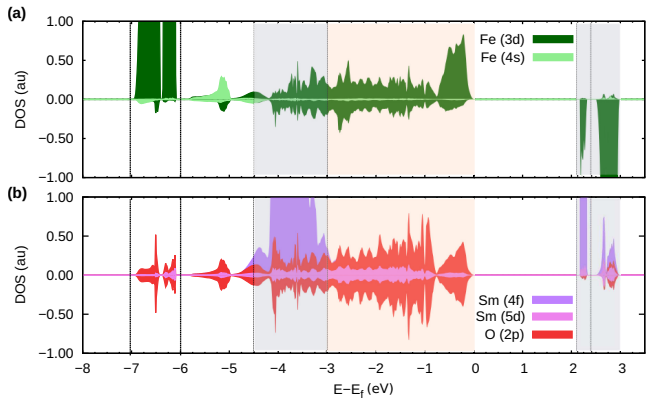

FIG. 2. Electronic density of state (DOS) for the antiferromagnetic **G**-type Sm and Fe spin order: filled $d$ and $f$ orbital states of Fe (green color) and Sm (violet color) are mainly in the intervals of (-7:-6) eV and (-4.5:-3) eV, respectively; $p$-states of O (red color) mainly occupy the top of the valence bands and hybridize with Fe-$d$ states. The conduction bands are characterized by empty Sm-$f$ and Fe-$d$ states. Vertical dashed lines guide visualization of the different energy regions; shadow areas highlight $p$-$d$ (O-Fe) and possible $d$-$f$ (Fe-Sm) hybridization regions. The related electronic band structure is shown in Fig. S1

In the SI we show the (predictable) behavior of the $4f$ and $3d$ states with the Hubbard $U$ parameters. No substantial changes in the hybridization or spin-polarization are observed.

## IV. MAGNETIC PROPERTIES: COLLINEAR MAGNETISM AND SOC EFFECT

As mentioned in the Sec. I, the Fe magnetic substructure of SFO experimentally undergoes a spin reorientation (SR) transition, i.e. the rotation of the magnetization vectors with respect to the crystallographic axis, between 450 K and 480 K. Upon decreasing temperature, the weak Fe-ferromagnetic (FM) moment rotates from the long $c$ axis to the short $a$ axis; the overall spin configuration thus changes from $\mathbf{G}_x A_y \mathrm{FM}_z$ to $\mathrm{FM}_x C_y \mathbf{G}_z$, with the AFM-**G**-type spin order remaining dominant. This is related to the fact that, in SFO as in the other $R\mathrm{FeO}_3$ compounds, the strongest magnetic exchange interactions are related to the Fe-substructure, and they determine the main magnetic ordering and *relative* orientation of the spins.

|  | HSE (meV) | Spin order | | MAE $\Delta E$ (meV/f.u.) |
|---|---|---|---|---|
|  |  | Sm | Fe |  |
| $J_1$ (Fe-Fe) | $\simeq 5.9$ | $\mathbf{FM}_z$ | $\mathbf{G}_z$ | $\simeq 0$ |
| $J_1$ (Sm-Sm) | $\simeq 0.0$ | $\mathbf{FM}_z$ | $\mathbf{G}_x$ | $\simeq 1$ |
| $J_2$ (Fe-Fe) | $\simeq 0.2$ | $\mathbf{FM}_x$ | $\mathbf{G}_z$ | $\simeq -44$ |
| $J_2$ (Sm-Sm) | $\simeq 0.0$ | $\mathbf{FM}_x$ | $\mathbf{G}_x$ | $\simeq -47$ |
| $J_1$ (Sm-Fe) | $\simeq 0.0$ | $\mathbf{C}_z$ | $\mathbf{G}_x$ | $\simeq -1$ |
|  |  | $\mathbf{C}_x$ | $\mathbf{G}_x$ | $\simeq -43$ |
|  |  | $\mathbf{C}_x$ | $\mathbf{G}_z$ | $\simeq -46$ |

TABLE II. (Left side) Heisenberg magnetic exchange (HSE) for first ($J_1$) and second ($J_2$) Fe-Fe and Sm-Sm neighbor interactions and Sm-Fe $J_1$ interaction, given in meV per atom (NB: the values are also divided by $S^2$, with $S = \frac{5}{2}$). (Right side) Magnetocrystalline anisotropy (MAE, in meV/f.u.): energy differences between various Sm and Fe magnetic orders taking into account the spatial orientation of the spins. A schematic of some of the magnetic structures is shown in Fig. 3.

The importance of the Fe sublattice in the magnetic ground state becomes clear in the DFT energies of magnetic configurations combining different spin orderings at the Sm- and the Fe- sites, as reported in Table SII: the AFM **G**-type ordering, is the ground-state for the Fe-spin substructure, independently of the magnetic ordering considered for Sm-spins. On the other hand, the Sm magnetic substructure shows a strong competition between ferromagnetic and antiferromagnetic configurations which are very close in energy.

In Table II, we report average Heisenberg spin exchange (HSE) estimated through an energy mapping method employing the DFT energy differences from six magnetic configurations among those listed in Table SII. This Fe AFM nature is clear from the strong first nearest neighbour exchange ($J_1$) of about 5.9 meV between Fe sites. Beyond the first nearest neighbour interaction, the exchange interactions fall rapidly to zero. In detail, we consider the Heisenberg spin Hamiltonian of the type $H = \frac{1}{2} \sum_{i \neq j} J_{ij} \mathbf{S}_i \cdot \mathbf{S}_j$, with $J_{ij}$ the isotropic Heisenberg exchange coupling between interacting spins $\mathbf{S}$ at the $i$ and $j$ sites; here we use $S = \frac{5}{2}$ for both Sm and Fe ions. We took into account six first- ($J_1$) and twelve second- ($J_2$) neighbors for the Fe-Fe and Sm-Sm magnetic interactions, and eight first-neighbor Sm-Fe interactions, treating all of the magnetic pairs as structurally equivalent, i.e. neglecting variations in the inter-atomic dis-tances related to the structural distortions described in the previous Sec. III A. As anticipated, the $J_1$ exchange coupling between Fe ions is strongly antiferromagnetic and is the leading term in the magnetic Hamiltonian of SFO; $J_2$ is one order of magnitude smaller. On the other hand, Sm-related interactions are on average very small (close to zero and within the DFT accuracy), in line with the quasi-degeneracy observed between the various FM and AFM Sm-spin orders.

Such very small Sm-Sm and Fe-Sm interactions reveals a strong frustration in the Sm spin lattice. Furthermore, the weak Sm-Fe exchange interaction shows more sensitivity than the Fe-Fe exchange to the spins order used for the spin Hamiltonian parametrization (to the Fe-spin order in particular). $J_1$(Sm-Fe) can vary from negative (FM) to positive (AFM) values, but remains a small contribution to the total energy, on the order of 0.1 meV. This leads to the approximately zero value reported in Table II. Similarly, the estimate of the weaker Sm-Sm interactions also depends on the magnetic configurations used for the energy mapping. In particular, a weakly antiferromagnetic $J$ (also of order 0.1 meV) is obtained when including only one **G**-type Fe- spin configuration among the six used, i.e. when performing the fitting with higher energy spin configurations instead of those nearer the ground-state.

Such fluctuations are not observed in the similarly small $J_2$ Fe-Fe interactions, which preserve magnitude and sign, remaining antiferromagnetic. These observations suggest that the weak Sm-spin interactions depend on the magnetic surroundings and that the interactions could be different, either FM or AFM, for each magnetic pair in the distorted SFO, giving rise to disorder and frustration in the Sm magnetic substructure [47, 48]. Likely supported by thermal fluctuations and entropic contributions, at finite temperature, such a behavior could justify the reported cluster glass state in the low-T Sm-spin ordering [10].

Noteworthy, despite the $U$-induced shifts of the Sm-$f$ and Fe-$d$ states in the VB and CB (Fig. S3), magnetic exchange energies (Fe exchange couplings, in particular) are not substantially affected by changing the Hubbard-$U$ correction on the Sm-$f$ and Fe-$d$ states, based on the tests we performed and reported in Tables SII-SIII. The weak influence of the Sm magnetic moment on the magnetic ordering and the relative Fe-spin orientation is further supported by calculations without the Sm-$f$ electrons. Freezing the Sm-$f$ electrons within the core states, and therefore removing the magnetic contribution of Sm, does not modify the magnetic interaction significantly (last column Table SII). This could be ascribed to the absence of strong $f$-$d$ hybridization which could support superexchange between the localized Sm-$4f$ spins; in turn, this is also in line with the almost insensitivity of the Fe and O electrons to the presence of explicit Sm-$f$ electrons (Sec. III B).

So far, we have discussed the isotropic magnetic exchange interactions, with results showing the much weaker $R$-exchange couplings with respect to the stronger and dominant AFM Fe-Fe interaction in SFO. We now look at the effect of the spin-orbit coupling in the magnetic interactions, the magnetocrystalline anisotropy (MAE) in particular.

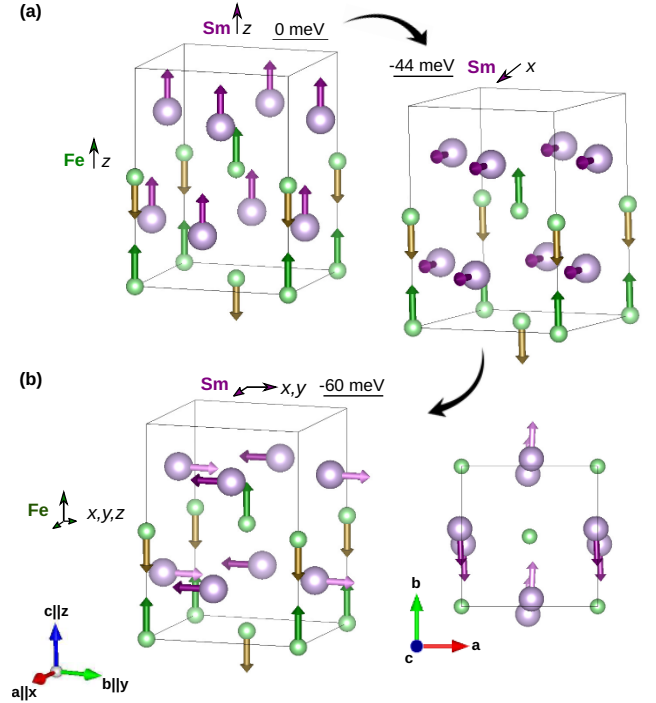

FIG. 3. Schematic view of **(a)** $\mathbf{FM}_z$-$\mathbf{G}_z$, $\mathbf{FM}_x$-$\mathbf{G}_z$ and **(b)** $\mathrm{FM}_x\mathbf{C}_y$-$\mathrm{FM}_x\mathrm{C}_y\mathbf{G}_z$ noncollinear Sm-Fe spin orders. Top view in **(b)** shows the little Sm-spin ferromagnetic canting. Energy gains (in meV/f.u.) from Table II and Table III are also reported.

In Table II, we report the energy differences between various spins states when including SOC in the DFT simulations. Particularly, we first looked at the energies associated to the **FM**-**G** and **C**-**G** configurations when Sm and Fe spins are: *i)* collinear and aligned along either the $x$ or $z$ direction, that provides an estimate of the total MAE in the crystal; *ii)* noncollinear, with Sm-spins aligned along the $x$ direction and the Fe-spins along the $z$ direction and *vice versa*, to estimate the MAE associated to the distinct magnetic substructures when comparing energies with respect to the previous configurations. Schematic examples are shown in Fig. 3a. We performed these calculations by constraining the magnetic moments along the wanted directions to avoid energetic contributions from additional spontaneous spin components.

SFO is characterized by a strong magnetic anisotropy driven by the Sm ions. In fact, all magnetic orders with Sm-spins aligned along the $x$-direction are lower in energy with respect to configurations with Sm-spins along the $z$-direction. At variance, Fe-spins experience a much lower energy cost by having spins along the $x$ or $z$ direction. We also checked the MAE, i.e. $\Delta E\,(\mathbf{G}_z\text{-}\mathbf{G}_x)$, associated to the Fe magnetic substructure when treating $f$-electron as core states, still obtaining a very small anisotropy, smaller than 1 meV/f.u.

Then, we removed the DFT magnetization constraint and let the electronic system relax to its energy minimum configuration, starting from $\mathbf{G}_z$ order for the Fe-spin substructure and either $\mathbf{FM}_x$ or $\mathbf{C}_y$ orders for the Sm spins. In both cases, the system spontaneously develops additional spin components both in the Fe- and Sm-spin substructure: Fe-spins develop little antiferromagnetic C-type ordering along the $y$-direction and weak ferromagnetism along the $x$-direction, which gives rise to a little spins canting and to the observed net magnetization; Sm-spins develop additional C-type and ferromagnetic components along the $y$ and $x$ directions, respectively, according to the two initial configurations. Calculated total energies and electrons moment components are reported in Table III. The $f$ states of Sm display an important orbital moment (OM) of opposite sign with respect to the magnetic moment (MM) associated with the $S=\frac{5}{2}$ spin state. Particularly, in the found lowest energy magnetic configuration, the $\mathrm{FM}_x\mathbf{C}_y\text{-}\mathrm{FM}_x\mathbf{C}_y\mathbf{G}_z$ (Sm-Fe) configuration (Fig. 3b), the OM and MM components associated to the ferromagnetic order along the $x$ direction are of the same order of magnitude, bringing to overall small moment and, in turn, to a Sm-$\mathrm{FM}_x$ moment competing with the Fe-$\mathrm{FM}_x$ moment. Such delicate balance can be rather sensitive to any kind of small changes, which can be related to the DFT approximations, such as $U$-correction or different exchange and correlation functionals, underlying atomic positions or crystal structure. Moreover, at present, it is not that straightforward to identify a clear ground state from DFT simulations, because of the presence of various and competing metastable spin states associated to different possible ways in occupying the $f$-orbital states of Sm, hence in defining the spin-density (occupation) matrix between the correlated orbitals.

Nonetheless, our simulations put forward two main evidences, so far only envisioned upon experimental observations: *i)* the major role played by the Sm-$4f$ electrons in the Fe-spin reorientation transition; we estimate a strong magnetic anisotropy associated to the Sm ions, favoring in-plane orientation of spins. *ii)* spontaneous development of canted-spin orders and non-collinearity between the main magnetic ordering of the Sm- and Fe-spin substructures, eventually favoring little and competing ferromagnetic moments along the $x$ direction (*or* the $\mathbf{a}$ crystallographic axis in the $Pbnm$ structure).

## V. LATTICE DYNAMICS

In order to investigate possible spin-phonon coupling effects, we calculated the phonon frequencies ($\omega$) under various conditions and compered these results with low-T Raman active vibrations reported by M. C. Weber and coauthors in Ref. [29]. Below the SR temperature, au-thors report two kinds of deviations from typical thermal behavior in the evolution of some phonon frequencies, $\omega(T)$, associated with the rare-earth and oxygen atoms displacements (see Fig. 3 in Ref, [29]): an increase in the $\omega(T)$ slope, *i.e.* phonons stiffening, of the $B_{2g}(1)$, $A_g(2)$ and $B_{3g}(2)$ modes; decreasing $\omega(T)$, *i.e.* phonons softening, for the $A_g(3)$ and $B_{1g}(3)$ modes. Spectral changes are evident at T$\simeq$ 200 K, where a possible metastable cluster glass phase is observed [10]. Similar phonon anomalies have also been observed for other oxide perovskite systems, such as BiFeO$_3$ [49] in the vicinity of $T_N$, GdFeO$_3$ at the $R$-spin ordering temperature [50] or TbMnO$_3$ [51] and also in different compounds such as BiMn$_3$Cr$_4$O$_{12}$ or CdCr$_2$S$_4$ associated with Cr-magnetism and the Cr-anion interactions [52, 53]. Such features are considered a signature of strong spin-phonon coupling.

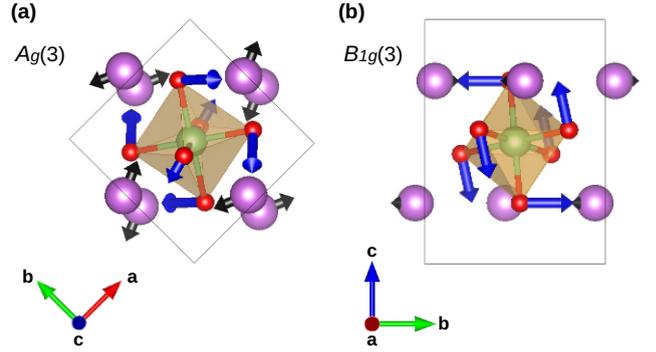

FIG. 4. Schematic view of the atomic pattern of distortion of the anomalous [29] $A_g(3)$ and $B_{1g}(3)$ phonon modes in the $Pbnm$ SFO structure, involving mainly: **(a)** antipolar motion of Sm-ions, FeO$_6$ octahedra tilting and rotation around the crystallographic **c**-axis; **(b)** octahedra rotation in the **b-c** ($y$-$z$) plane, respectively.

In Table IV, we report phonon frequencies for the Raman active modes calculated at the center ($\Gamma$) of the $Pbnm$-Brillouin zone. The 60 phonon modes comprises: $B_{1u}+B_{2u}+B_{3u}$ acoustic modes, $8A_u$ silent modes, $7A_g+7B_{1g}+5B_{2g}+5B_{3g}$ Raman (RM) active modes, $7B_{1u}+9B_{2u}+9B_{3u}$ infrared (IR) active modes [decomposition in the $Pnma$ setting is : $7A_g+5B_{1g}+7B_{2g}+5B_{3g}$ RM, $9B_{1u}+7B_{2u}+9B_{3u}$ IR [54]. Symmetry labels transformation from the $Pnma$ setting to the $Pbnm$ is: $B_{1g(u)} \rightarrow B_{3g(u)}$, $B_{2g(u)} \rightarrow B_{1g(u)}$, $B_{3g(u)} \rightarrow B_{2g(u)}$.].

We considered the relaxed atomic structure and lattice obtained either with or without the Sm-$f$ electrons in the valence states (*cfr.* Table I). For both structures we calculated vibrational energies w/ and w/o the Sm-$f$ states. We also considered the third structure obtained by optimizing the crystal structure with the $f$ electrons but applying a hydrostatic pressure of about 1 GPa, in order to distinguish volume effects from anisotropic strain effects (*cfr.* Table I). In this way, we can fully disentangle structural from Sm-$f$-induced effects on the observed spectra.

Interestingly, the five reported anomalous modes

| Sm-order | Fe-order | $\Delta E$ (meV/f.u.) | | Sm ($M_x$) | Sm ($M_y$) | Sm ($M_z$) | Fe ($M_x$) | Fe ($M_y$) | Fe ($M_z$) |
|---|---|---|---|---|---|---|---|---|---|
| **FM$_x$C$_y$** | FM$_x$C$_y$**G$_z$** | $\simeq$ -57 | **MM** ($\mu_B$) | 4.85 | -1.07 | 0.00 | -0.01 | -0.02 | 4.14 |
| | | | **OM** ($\mu_B$) | -2.98 | 0.92 | 0.00 | 0.00 | 0.00 | 0.01 |
| FM$_x$**C$_y$** | FM$_x$C$_y$**G$_z$** | $\simeq$ -60 | **MM** ($\mu_B$) | 0.77 | -4.91 | 0.00 | 0.03 | -0.02 | 4.14 |
| | | | **OM** ($\mu_B$) | -0.67 | 3.03 | 0.00 | 0.00 | 0.00 | 0.01 |

TABLE III. Noncollinear canted spin states with main ferromagnetic spin alignment along $x$-direction (**FM**$_x$C$_y$) or main C-type antiferromagnetic order along $y$-direction (FM$_x$**C**$_y$) for Sm. Bold characters highlight the main spin order for the two, Sm and Fe, magnetic substructures. Magnetic (MM) and orbital (OM) components of the spin moment for both Sm and Fe are reported. The reference for the energy difference is the **FM**$_z$-**G**$_z$ spin order of Table II.

| | w/o Sm-$f$ structure | | w/ Sm-$f$ structure | | 1 GPa structure | |
|---|---|---|---|---|---|---|
| **Symmetry** | **Sm-$f$ core** | **Sm-$f$ valence** | **Sm-$f$ core** | **Sm-$f$ valence** | **Sm-$f$ valence** | **Exp.[29]** |
| A$_g$(1) | 110 | 112 | 104 | 107 | 107 | 110* |
| B$_{1g}$(1) | 110 | 114 | 106 | 110 | 111 | 110 |
| **B$_{2g}$(1)** | **134** | **131** | **127** | **125** | **127** | **149** |
| **A$_g$(2)** | **139** | **145** | **127** | **134** | **136** | **146** |
| B$_{3g}$(1) | 150 | 150 | 152 | 151 | 152 | 161 |
| B$_{1g}$(2) | 160 | 162 | 149 | 152 | 154 | 158 |
| **B$_{3g}$(2)** | **236** | **244** | **210** | **218** | **220** | **241** |
| **A$_g$(3)** | **247** | **245** | **232** | **229** | **231** | **220** |
| **B$_{1g}$(3)** | **281** | **277** | **255** | **252** | **254** | **253** |
| B$_{2g}$(2) | 313 | 314 | 306 | 306 | 308 | 323 |
| A$_g$(4) | 319 | 322 | 306 | 310 | 312 | 319 |
| B$_{1g}$(4) | 345 | 345 | 344 | 342 | 345 | – |
| B$_{3g}$(3) | 354 | 353 | 356 | 353 | 357 | 354 |
| A$_g$(5) | 386 | 389 | 374 | 377 | 380 | 380 |
| A$_g$(6) | 417 | 415 | 420 | 419 | 424 | 421 |
| B$_{3g}$(4) | 426 | 424 | 427 | 426 | 432 | 426 |
| B$_{2g}$(3) | 425 | 421 | 433 | 429 | 435 | 433 |
| B$_{2g}$(4) | 449 | 445 | 446 | 443 | 446 | 456 |
| **B$_{1g}$(5)** | **460** | **460** | **446** | **446** | **449** | **463** |
| **A$_g$(7)** | **468** | **466** | **459** | **457** | **461** | **470** |
| B$_{1g}$(6) | 512 | 516 | 503 | 505 | 508 | 521 |
| B$_{3g}$(5) | 597 | 593 | 604 | 560 | 605 | – |
| B$_{1g}$(7) | 619 | 616 | 620 | 617 | 622 | 640 |
| B$_{2g}$(5) | 648 | 645 | 652 | 650 | 655 | – |

TABLE IV. Phonon frequencies (in cm$^{-1}$) of the Raman active modes in the *Pbnm* crystal structure, considering collinear G-type AFM magnetic ordering for the Fe atoms w/o Sm-$f$, and for both Sm and Fe spins sub-structures w/ Sm-$f$. In order, underlying atomic structures are fixed to: the one optimized w/o Sm-$f$; the one optimized w/ Sm-$f$ and collinear **G**-type Sm and Fe spins order; to the one optimized taking into account $f$ states and further hydrostatic pressure of about 1 GPa (*cfr.* Table I). Experimental data from Raman spectroscopy performed at 4 K have been provided by authors of Ref. [29]; * at 80 K taken from Ref. [27]

(violet-row in Table IV) exhibit strong coupling with the underlying crystal structure: frequencies of the B$_{2g}$(1), A$_g$(2) and B$_{3g}$(2) modes (which stiffen) are better reproduced when considering structural distortions in the w/o Sm-$f$ structure. Frequencies of the A$_g$(3) and B$_{1g}$(3) modes (which soften) are well reproduced in the w/ Sm-$f$ structure. Moreover a direct spin-phonon coupling effect is observed for the A$_g$(2) and A$_g$(3) modes, with a further hardening and softening, respectively, induced by the magnetic Sm atoms. Both the A$_g$(2) and A$_g$(3) modes involve in-plane Sm-antipolar motion, with additional in-phase octahedra rotations and tiltings for the A$_g$(3) mode (Fig. 4a). The involvement of Sm displacements could justify a direct coupling between the vibration and the Sm spins. Moreover, we also observe a possible interesting coupling with strain. In the w/o Sm-$f$ structure, the ***a*** structural parameter is smaller than in the w/ Sm-$f$ structure, producing a hardening of these modes; at variance, the reduced volume alone does not substantially affect the frequency. Similarly, also the B$_{1g}$(3) mode displays a coupling to strain, but weaker direct spin-phonon coupling than the A$_g$(3) mode. The B$_{1g}$(3) mode is in fact characterized by in-phase rotations but smaller Sm distortions than the two A$_g$ modes

(Fig. 4b).

The interpretation of the behavior of the $B_{2g}(1)$ and $B_{3g}(2)$ modes is less direct: both modes display some strain-phonon coupling, but weak spin-phonon coupling, despite the fact that their distortion patterns involve motion of the Sm ions (plus out-of-phase oxygen rotations for $B_{3g}(2)$). The Sm motion in these two modes occurs along the $c$ crystallographic axis, which suggests a relation to the anisotropic character of the Sm-related magnetic interactions.

The remaining Raman active modes are less sensitive to the structural distortions and coupling with Sm-$f$ states (within 10 cm$^{-1}$), and they are in agreement with the low-T data. Noteworthy, the measured vibration frequencies of the $B_{1g}(5)$ and $A_g(7)$ (grey-row in Table IV) modes are the only ones which match better with calculations performed in the w/o Sm-$f$ structure.

In order to investigate coupling of phonons with SOC and Sm-magnetism, in Table SIV, we report frequencies of the Raman active modes for different Sm-magnetic orders. Collinear antiferromagnetic or ferromagnetic Sm-spin orders do not affect phonons, whereas variations between 3 and 5 cm$^{-1}$ are induced on the anomalous $B_{2g}(1)$, $B_{3g}(2)$ and $A_g(3)$ modes by SOC effects, when considering the non-collinear $FM_x\mathbf{C}_y$-$FM_xC_y\mathbf{G}_z$ (Sm-Fe) configuration (Table III). Such effects are almost absent when excluding Sm-$f$ electrons.

Additionally, in Table SV, we report the silent and infrared (IR) modes of the $Pbnm$ structure. Unfortunately, no IR experimental spectra are available. Interestingly, the silent $A_u(1)$ mode and the $B_{3u}(2)$, $B_{1u}(2)$, $B_{1u}(4)$ and $B_{1u}(5)$ IR modes show sensitivity to the Sm-$f$ states, whereas the $B_{3u}(1)$, $B_{1u}(1)$, $A_u(4)$, $B_{3u}(4)$, $B_{2u}(4)$, $B_{3u}(5)$, $B_{2u}(6)$, $B_{1u}(5)$ and $B_{3u}(8)$ show coupling to strain.

## VI. DISCUSSION AND CONCLUSION

By means of first-principles calculations, we have investigated structural, electronic, magnetic and dynamical properties of SmFeO$_3$ unveiling spin-spin, spin-lattice and spin-phonon couplings effects driven by the magnetic Sm ion. In particular, we performed comparative simulations with and without inclusion of the $f$ electrons in the valence states, with the further inclusion of the spin-orbit coupling. The energy mapping over various Sm and Fe spins orders revealed a magnetic frustration of the Sm spin lattice due to very weak and eventually competing ferromagnetic and antiferromagnetic Sm-Sm and Sm-Fe exchange interactions against robust antiferromagnetic Fe-Fe exchange interactions. This supports the lack of Sm magnetic ordering at high-temperature and possible disordered spin states at medium temperature. At very low-temperature, the reduction of the energy cost related to thermal fluctuations and entropy could allow for the Sm spin moments ordering, stabilized by a large magneto-crystalline anisotropy. In fact, we found out that the Sm spins largely prefer to lay in the $\{a,b\}$ plane of the $Pbnm$ crystal structure, perpendicularly to the Fe spins, due to a strong spin-lattice coupling mediated by the SOC of the Sm ions created by the $4f$ electrons. We thus identified the large anisotropy of the Sm magnetic ions as a possible microscopic mechanism driving the experimentally observed Fe spin reorientation transition in SmFeO$_3$. Furthermore, the small canting of both Sm and Fe spins associated with the non-collinear ground-state, gives rise to a weak ferromagnetic moment along the $a$-axis. Interestingly, we found that Sm spins exhibit large orbital moment of opposite sign with respect to the magnetic moment, which could be the origin of the magnetization reversal, hence anti-parallel Sm and Fe spins, observed at low-T. Additional evidences of the magnetic activity of the Sm-$f$ electrons are also observed in the lattice vibrations. We pointed out in fact a direct coupling between specific phonon modes and Sm spins, and also further indirect evidences of the anisotropic character of the Sm-$f$ electronic interactions, which are in line with experimentally reported anomalous vibrations. Furthermore, we also put forward interesting phonon-strain couplings.

This work gives a deep insight into the underlying physics of the rich SmFeO$_3$ phase diagram and provide key ingredients for future works, both from the theoretical side, to investigate, for instance, finite-temperature and external perturbations effects through Monte Carlo and/or spin dynamics simulations, and from the experimental side to improve low-T characterization of the structural and magnetic properties.

## ACKNOWLEDGMENTS

This work was supported by the m-era.net project SWIPE funded by FNRS Belgium grant PINT-MULTI R.8013.20, and by computing time on the LUMI supercomputer and on the Tier-1 supercomputer of the Fédération Wallonie-Bruxelles infrastructure (grant agreement n. 1117545), through the Consortium des Équipements de Calcul Intensif. The authors acknowledge M.C. Weber and the SWIPE project partners, in particular M. Guennou, G. Gordeev, M. Bibes, and L. Iglesias, for insightful discussions and scientific exchanges. D.A. acknowledges E. Bousquet and A. Sasani for some technical support and implementation of the routine for the orbital moment (OM) calculation in the ABINIT package (release and publication in preparation).